\documentclass[onecolumn,amsmath,superscriptaddress,amssymb]{revtex4}

\usepackage{graphicx}

\begin{document}

\title{On the Possibility of Medium-Energy Compact X-ray Free-Electron Laser}

\author{Lekdar Gevorgian}
\affiliation{Alikhanian National Scientific Laboratory (Yerevan Physics Institute) foundation, Theory Division - Alikhanian Brothers str. 2, 0036, Yerevan, Armenia}
\author{Valeri Vardanyan}
\email{vardanyanv@yerphi.am}
\affiliation{Alikhanian National Scientific Laboratory (Yerevan Physics Institute) foundation, Theory Division - Alikhanian Brothers str. 2, 0036, Yerevan, Armenia}
\affiliation{Yerevan State University, Faculty of Physics - Alex Manoogian str. 1, 0025, Yerevan, Armenia}

\begin{abstract}

The problem of X-ray Free-Electron Laser operating on self-amplified spontaneous emission in irregular microundulator is considered. The case when the spectrum width of spontaneous radiation is conditioned by the spatial distribution of sources creating the undulating field is considered. In this case gain function of the stimulated radiation is dozens of times higher than that of the conventional undulators. We propose a model of irregular microundulator, which can be used to construct a drastically cheap and compact X-ray free-electron laser operating on medium energy electron bunch.

\end{abstract}


\maketitle

\section{Introduction}

Generation of powerful photon beams in different frequency ranges due to its wide field of applications in physics and other science disciplines such as biology, chemistry, material science, etc. is one of the biggest challenges in modern physics. For that purpose alongside with the traditional lasers so-called free-electron lasers (FEL) are also being widely used due to their well known features such us tunability on every required wavelength. The idea to use FELs to generate powerful photon beams \cite{Madey} was first realized in amplification regime during the experiment \cite{Elias} and in generation regime during the experiment \citep{Deacon}. Photon beam with 109~nm wavelength was generated at Tesla test facility for free-electron laser experiments (TTF FEL) at DESY using electron bunch with 233 MeV energy and operating on self-amplified spontaneous emission (SASE) \cite{Andruszkow}. Photon beam with a 1.5~\AA wavelength was generated using 13.6 GeV energy electron bunch of the Stanford linear accelerator center's linac coherent light source X-ray FEL (SLAC LCLS XFEL) again using SASE \cite{Emma}.

Even though modern FELs solve the problems posted in front of them, they are still very big and expansive devices. The reason is that to create necessary for the applications short wavelengths one must either put the undulator magnets close enough to each other, which still is a big practical challenge, or to use high-energy electron bunches (see, for example, \cite{Saldin_book} or \cite{Marshall}).

As a result we have FELs with very big sizes and very high maintenance costs. Of course there are steps towards the development of FEL which will be compact in its sizes and will operate with low-energy electron bunches. As an example one can mention the letter \citep{Dai}, where the authors propose a FEL which is based on the higher harmonics of the spontaneous radiation of the 3.5 GeV energy electron bunch in FEL oscillator.

In this paper we announce about the possibility of an efficient (i.e. cheap and compact compared with the existing ones) FEL based on an undulator with randomly varying spatial period which is in order of several micrometers. Hereafter we will refer to such an undulator as "irregular microundulator". One can imagine the irregular microundulator as a set of charged needles placed parallel to each other with random distances between two adjacent needles in order of few micrometers. The electron bunch, being shot in direction perpendicular to the needles into the inter-needle space, will be repealed by the charged needles and perform an irregular undulatory motion. We will show that the practical realization of such a microundulator will allow to construct significantly cheap and compact FEL.

In the past the problem of electron radiation moving along irregular periodic paths was solved in the work \citep{Gevorgian_1}, where the radiation from non-relativistic electrons conditioned by the interactions with randomly distributed roughness of the metallic surface was considered. A Smith-Purcell type radiation \citep{Smith} spectrum was obtained which allowed to explain the anomalously high intensity in the transition radiation experiments \cite{Blanckenhagen, Jones} from non-relativistic electrons grazing into metallic surfaces.

\section{The physical model}

In this problem one should choose the mean transversal distance between adjacent needles to be in order of magnitude of the mean distance between electrons in the bunch. The longitudinal distance should be chosen from the requirement to obtain photon beam with the given wavelength. The needles should be many times longer than the transversal size of the bunch.

\begin{figure}[h]
\begin{center}
\includegraphics[scale=0.5]{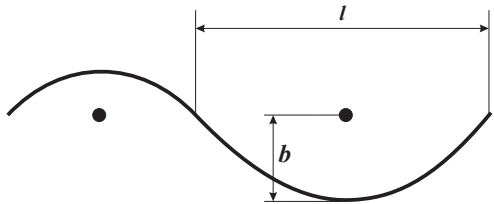}
\end{center}
\vspace{-10pt}
\caption{The path of the electron}
\vspace{-5pt}
\label{path}
\end{figure}

In such a microundulator bunch electrons propagating in $z$ direction, being repealed from the charged needles will oscillate in the $xz$ plane in the $x$ direction. Let us assume the electron trajectories consisting of sinusoidal curves with amplitude $b$ and spatial semiperiod $l$: $x(z) = b \sin(\pi z/l)$. Such a trajectory is presented in Fig.~\ref{path}. From the condition of trajectory smoothness it follows that $\pi b/l = \beta_\perp/\beta_\parallel = const$, where $\beta_\perp c$ is the maximal speed of the electron in the $x$ direction, $c$ is the speed of light in vacuum, $\beta_\parallel c$ being the mean speed in the $z$ direction. Because the oscillations in the $x$ direction are accompanied by the oscillations in the $z$ direction, for the mean square root of $\beta_\parallel$ one will get $\langle \beta_\parallel^2 \rangle = \beta^2 - \beta_\perp^2/2$, where $\beta$ is assumed to be constant (we neglect the small energy losses conditioned by the radiation). From here it also follows that the undulator parameter $q = \beta_\perp \gamma/\beta_\parallel$ of the electrons moving along trajectories with random spatial semiperiods is a constant quantity ($\gamma = (1-\beta^2)^{-1/2}$ is the Lorentz-factor). Hence for the Lorentz-factor conditioned by the longitudinal motion one can get: $\gamma_\parallel^2 = (1-\beta_\parallel^2)^{-1/2} = \gamma^2/Q$, where $Q = 1+q^2/2$.

Since the bunch electrons are indistinguishable from each other one can assume that each electron moves not along a trajectory which is made up from the embroidering of sinusoidal trajectories with different spatial semiperiods, but along a sinusoidal trajectory with unique spatial semiperiod $l$. However, this assumption implies that the different electrons move along trajectories with different spatial semiperiods. As a consequence of the finiteness of the number of electrons in the bunch, $l$ is being a discrete quantity, but in our problem this discrete quantity can be considered as a continuum quantity, which obeys to the gamma distribution law. It is convenient to use the gamma distribution for the $t = l/\langle l \rangle$ parameter, where $\langle l \rangle$ is the mean value of the spatial semiperiods $l$. This distribution is presented in the Eqn.~(\ref{gamma}):

\begin{equation}\label{gamma}
f(a,t) = \frac{a^at^{a-1}e^{-at}}{\Gamma(a)},\;   a>0.
\end{equation}

\noindent
Here the scaling parameter of the distribution is taken to be equal to the parameter $a$ which is the dispersion of the distribution and is conditioned by the non-regularity of the spatial semiperiod distribution. Since $\langle t \rangle = 1$ the degree of non-regularity $\eta$ is being determined from the parameter $a$: $\eta = \sqrt{\langle \Delta t^2 \rangle} = 1/\sqrt{a}$.

\begin{figure}[h]
\begin{center}
\includegraphics[scale=0.8]{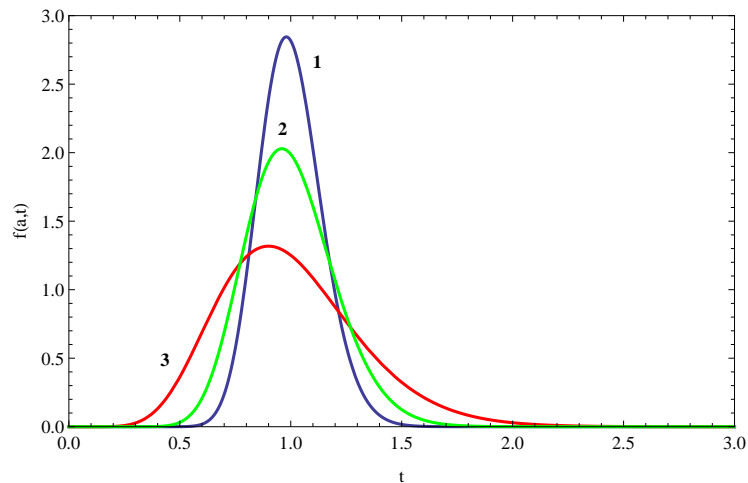}
\end{center}
\vspace{-20pt}
\caption{Gamma distribution for different values of $a$: 1) $a=50$, 2) $a=25$, 3) $a=10$.}
\vspace{-5pt}
\label{gamma_fig}
\end{figure}

In Fig.~\ref{gamma_fig} gamma distribution curves for different values of $a$ are presented.

\noindent
\section{Radiation characteristics} The radiation characteristics of the electron bunch are obtained by averaging the corresponding characteristics of the single electron radiation.

It is known that the radiation spectrum width is conditioned by the energetic and angular divergence of the bunch electrons, as well as by the finiteness and non-regularity of the electron motion. In the most of the cases the angular divergence is negligible in comparison with the energetic divergence. In such a case the radiation spectrum width is being calculated from Eqn.~(\ref{bandwidth}) \cite{Gevorgian_2}: 

\begin{equation}\label{bandwidth}
\frac{\Delta \omega}{\omega} \approx \frac{2l}{L}+[2(\frac{\Delta \gamma}{\gamma} + \eta)]^{1/2},
\end{equation}

\noindent
where $L$ is the length of the metallic plate.

Let us consider the case when the spectrum width is mainly conditioned by the random distribution of the quantity $l$ ($\eta\gg\Delta \gamma/\gamma, 2l/L$).

We should note that during the radiation process of the relativistic electron the energy and momentum conservation laws allow radiation only under the small angles ($\theta \lesssim 1/\gamma\ll1$). In such a case for the angular-frequency distribution of the number of photons radiated from a bunch electron which moves along a trajectory with a spatial period $t$, in the dipole approximation ($q = \beta_\perp \gamma<1$), switching to $u = \theta\gamma$ angles and $X = \omega \langle l \rangle / \pi c \gamma^2$ dimensionless frequencies we can get: 

\begin{equation}\label{freq_ang_dist}
\frac{d^2N}{dXdu^2} = BX[1+(1-tXu^2)^2]\delta(X(Q+u^2)-\frac{2}{t}),
\end{equation}

\noindent
where $B = \pi \alpha n q^2/8$, $\alpha$ is the fine structure constant and $n = L/ \langle l \rangle$ is the doubled number of electron oscillations along the distance $L$. 

To obtain the frequency distribution of the radiated photons one needs to integrate the Eqn.~(\ref{freq_ang_dist}) with respect to $u^2$. After the integration one has the Eqn.~(\ref{freq_dist}).

\begin{equation}\label{freq_dist}
\frac{dN}{dX} = B[1+(1-QXt)^2].
\end{equation}

\begin{figure}[h]
\begin{center}
\includegraphics[scale=0.9]{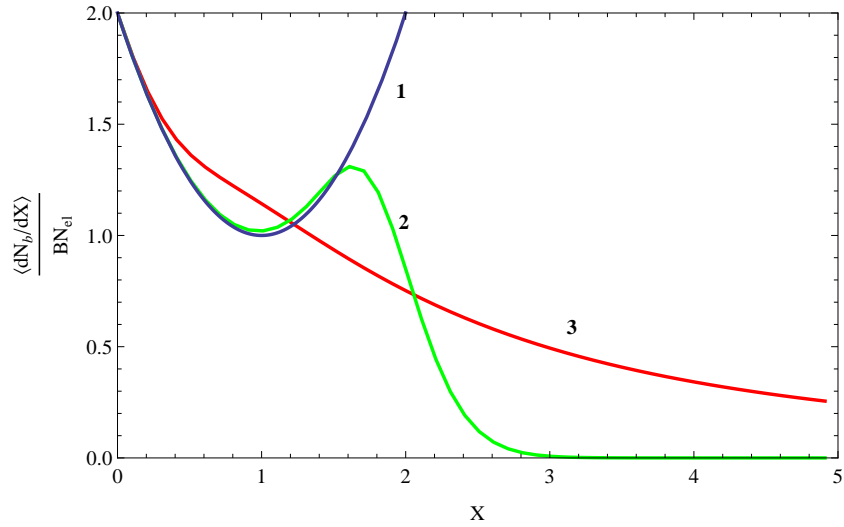}
\end{center}
\vspace{-20pt}
\caption{Frequency distribution of the bunch radiation for the regular case and for different values of $a$: 1) Eqn.~(\ref{freq_dist}), 2) $a=50$, 3) $a=2$.}
\vspace{-5pt}
\label{freq_bunch}
\end{figure}

To obtain the frequency distribution from the electron bunch one needs to average the Eqn.~(\ref{freq_dist}) by multiplying it by the gamma distribution given in Eqn.~(\ref{gamma}). Since the argument of the delta-function in Eqn.~(\ref{freq_dist}) should has a possibility to vanish, we can get a relation between $t$ and $X$: $t\leq2/X$. This is natural since it states that the harder photons are radiated by the electrons moving along trajectories with smaller spatial periods. Using this fact one gets for the frequency distribution of the radiation from the bunch:

\begin{equation}\label{freq_dist_bunch}
\left\langle \frac{dN_b}{dX}\right \rangle = BN_{el}\int_0^{2/QX}{f(a, t)[1+(1-QXt)^2]}{dt},
\end{equation}

\noindent
where $N_{el}$ is the number of electrons in the electron bunch.
In Fig.~\ref{freq_bunch} the bunch radiation frequency distribution curves are presented for different values of $a$ (the parameter $Q$ is taken to be equal to unity).

As it is easy to conclude from Fig.~\ref{freq_bunch}, at the hard frequency region the spectrum drastically differs from the regular case.

Let us mention that Eqn.~(\ref{freq_ang_dist}) in case of the regular motion ($t = 1$) coincides with Eqn.~(11) of the work \citep{Gevorgian_3} in the case of vacuum ($\sigma = 1$), if we take $Q = 1$. One can also mention that instead of Eqn.~(14.114) of the book \citep{Jackson} it is more precise to use Eqn.~(\ref{freq_ang_dist}).

A topic of special interest is the frequency distribution of the number of photons radiated under the zero angle which uniquely defines the gain of the stimulated radiation.

Substituting $u = 0$ in Eqn.~(\ref{freq_ang_dist}) one obtains Eqn.~(\ref{freq_dist_zero}) for the frequency distribution of the number of photons radiated under the zero angle from a single electron, which moves along a trajectory with a spatial period $t$:

\begin{equation}\label{freq_dist_zero}
\left.\frac{dN}{dX}\right\vert_{u = 0} = \frac{4B}{Q^2X}\delta(t-\frac{2}{QX}).
\end{equation}

For the frequency distribution of the number of photons radiated under the zero angle from the electron bunch we have:

\begin{eqnarray}\label{freq_dist_bunch_zero} \nonumber
\left\langle \frac{dN_b}{dX}\right \rangle_{u = 0} &=& \frac{2BN_{el}}{Q\Gamma(a)} F(a,X),\\
F(a,X) &=& Y^ae^{-Y}/\Gamma(a),Y = 2a/QX.
\end{eqnarray}

\begin{figure}[h]
\begin{center}
\includegraphics[scale=0.9]{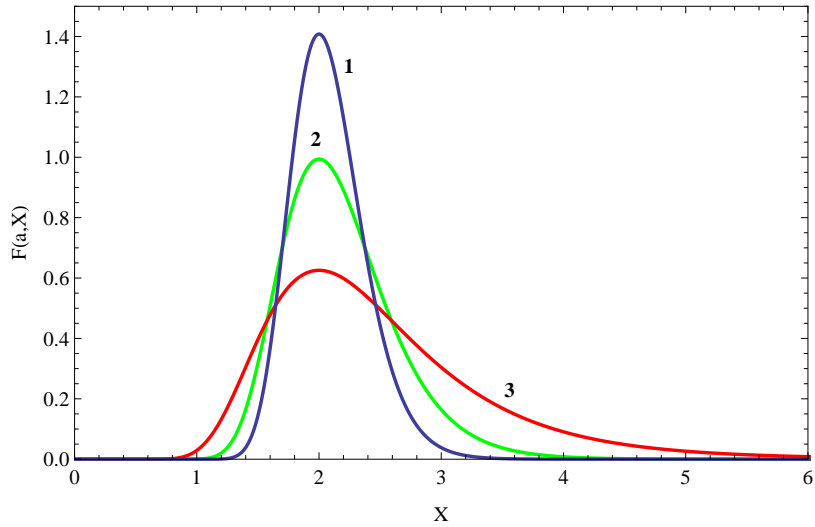}
\end{center}
\vspace{-20pt}
\caption{Frequency distribution under the zero angle for different values of $a$: 1) $a=50$, 2) $a=25$, 3) $a=10$.}
\vspace{-5pt}
\label{freq}
\end{figure}

In Fig.~\ref{freq} the zero-angle bunch radiation frequency distribution curves are presented for different values of $a$ (the parameter $Q$ is taken to be equal to 1).

\newpage
\section{X-ray SASE FEL in irregular microundulator} Following the work \citep{Gevorgian_4} for the linear gain from the interaction length $L$ of the stimulated radiation of the electron bunch with a electron density $\rho$ we derive:

\begin{eqnarray}\label{gain}\nonumber
G(a,X) = G_0\times g(a,X),\; G_0 &=& \frac{(2 \pi)^3q^2L^2r_0\rho}{\gamma^3},\\
g(a,X) &=& -X\frac{\partial F(a,X)}{\partial X},
\end{eqnarray}

\noindent
where $r_0 = 2.8\times 10^{-13}$~cm is the classical radius of electron.

\begin{figure}[h]
\begin{center}
\includegraphics[scale=0.7]{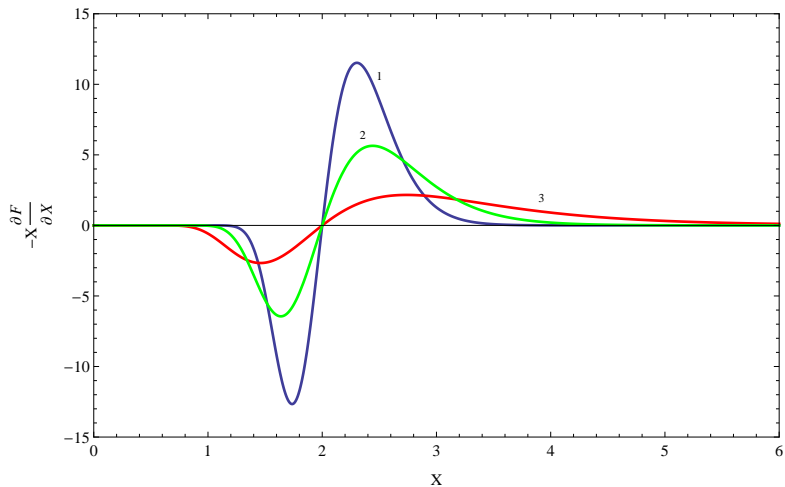}
\end{center}
\vspace{-20pt}
\caption{Function $g(a, X)$ for different values of $a$: 1) $a=50$, 2) $a=25$, 3) $a=10$.}
\vspace{-5pt}
\label{derivative}
\end{figure}

So, the gain in case of a non-regular microundulator is defined by the function $g(a,X)$. The plot of this function for different values of $a$ is shown in Fig.~\ref{derivative}. Let us mention that the function $g(a,X)$ for realistic values of $a$ is in order of magnitude higher compared with the regular undulator case \citep{Marshall}.

Taking into account the rapid growth of the nanotube technology we propose carbon nanotubes as repealing needles for the discussed microundulator. Carbon nanotube is a huge molecule of carbon rolled into a cylinder with a radius of few nanometers and with a macroscopic length \cite{Nano_1, Nano_2}. We hope that in the near future it will be possible to develop an irregular nanotube super-lattice placed between two parallel charged metallic surfaces, which can serve as a microundulator. The reason of such a hope is already existing technology of producing nanotube forest \cite{Nano_3}. 

\section{Discussion}

Now let us calculate the gain for the parameters of the TESLA Test Facility for FEL Experiments \cite{Andruszkow}. As a microundulator for the FEL we propose to use a super-lattice which is made of nanotubes, where the mean distance between adjacent nanotubes in direction of the bunch motion is in order of $\langle l \rangle = \lambda\gamma^2X_{\rm max}/2$, where $\lambda$ is the wavelength of the radiated photon and at $X_{\rm max}$ the function $g(a,X)$ achieves its maximal value. From Fig~\ref{derivative} it is seen that $X_{\rm max} = 2.3$. The mean distance between adjacent nanotubes in the transversal direction is in order of $\rho^{-1/3}$. The energy of the bunch is $E = 233$~MeV ($\gamma \approx 4.6 \times 10^2$), the number of electrons in the bunch is $10^{10}$, the transversal size of the bunch is $10^{-2}$~cm (therefore the nanotubes should be $10^{-1}\div 1$~cm long). Taking for the longitudinal size of the bunch $\sim 10^{-2}$~cm, for the electron density we get $\rho = 10^{16}\;{\rm cm}^{-3}$. If $\langle l \rangle = 3.6 \times 10^{-3}$~cm ($\lambda = 1.5 \AA$) and the electric field amplitude near the nanotubes is $F = 5.5\times 10^5$~CGSE units, then for the undulator parameter one gets $q \approx 2\times 10^{-4}\langle l\rangle F \approx 0.4$. For the irregularity of $\eta \approx 0.14$ ($a = 50$) the function $g(a,X)$ achieves its maximum value, $g(a,X)\approx 11.5$, at the frequency $X_{\rm max} = 2.3$.

As can be noted from Eqn.~(\ref{derivative}), for the photons of energy $8.3$~KeV ($\lambda = 1.5 \AA$) the linear gain from the $L \gtrsim 10$~cm length of the path takes a value bigger than 1. For smaller values of the electric field around the nanotube needles, the gain can be increased by choosing longer plates.
\newpage
\section{Conclusion}

We propose a compact X-ray Free-Electron Laser (CXFEL) operating on SASE regime, which uses medium-energy electron accelerator and few centimeter long non-regular microundulator. In case of overcoming the technical difficulties of creation of such a microundulator it is possible to generate 8.3~KeV energy photon beam from 10~cm length of the interaction path, using TTF FEL's 233~MeV energy electron bunch instead of SLAC's 13.6 GeV energy electron bunch.

\end{document}